\documentclass[twocolumn,widetext,showpacs,preprintnumbers,amsmath,amssymb]{revtex4}
\usepackage{graphicx}
\usepackage{dcolumn}
\usepackage{bm}

\begin{document}
\title{Extracting directed information flow networks: an application to genetics and semantics}
\author{A.P. Masucci$^1$}
\author{A. Kalampokis$^2$}
\author{V.M. Egu\'iluz$^1$}
\author{E. Hern\'andez-Garc\'ia$^1$}
\affiliation{
1-Instituto de F\'{\i}sica Interdisciplinar y
Sistemas Complejos IFISC (CSIC-UIB), E-07122 Palma de Mallorca,Spain
}
\affiliation{
2-Department of Marine Sciences, University of the Aegean, 811 00 Mytilene, Lesvos, Greece}

\date{\today}

\begin{abstract}
We introduce a  general method to infer the directional information flow between populations whose elements are described by $n$-dimensional vectors of symbolic attributes. The method is based on the Jensen-Shannon divergence and on the Shannon entropy and has a wide range of application. We show here the results of two  applications: first extracting the network of genetic flow between the meadows of the seagrass Poseidonia Oceanica, where the meadow elements are specified by sets of microsatellite markers, then we extract the semantic flow network from a set of Wikipedia pages, showing the semantic channels between different areas of knowledge.
\end{abstract}
\maketitle

\section{Introduction}
Since recent years \cite{43,49} network theory has become a hot topic among scientists. Its applicability to many fields of science made of network theory a primary tool to understand a wide range of phenomena, both in pure science  and in policy management \cite{a2}. In particular networks can be used whenever we want to understand the morphology and the topology of a complex system of interacting elements. Such a topology can then be useful to predict and to understand the complex behaviours of those systems. This kind of analysis is particularly relevant in the case of the study of infection spreading, both in society and in the WWW \cite{20}, or for understanding the formation and modification of urban conglomerates \cite{21}.

Most of the times the vertices of a network are single elements, such as individuals in social networks, tokens in linguistic networks, proteins in metabolic networks \cite{40}, etc.
 Nevertheless we can consider networks where the vertices are not single elements, but ensembles of elements, such as populations of a metapopulation \cite{a11,a5}, and the links are given by some relations between those ensembles.

 In particular suppose that we have a set of $N$ populations, where each element of a population is described by a $n$-dimensional vector of symbolic or numeric attributes. Those attributes  can be a set of social or ethnical indicators in the case of social systems (age, qualification level, salary, etc.), a set of genetic markers for biological populations, etc. Then  each population can be represented by its probability distribution in the $n$-dimensional attribute space and by its size, where the probability distribution counts the probability that an element in a population is characterised by a given vector in the attribute space.
Thus we can consider the network where each population is a vertex and two vertices are linked whenever an \emph{information flow} between the two relative probability distributions is detected.

We use the term information flow, instead of \emph{correlation} or \emph{distance}, because we want to emphasise that in the systems we consider often the correlations arise from  migration or inheritance of elements between different populations. Then the movement of elements from a population to another one corresponds to a movement of attributes or a flow of information in the attribute space where those populations are defined.

In general we can say that there is an information flow between two
attribute distributions if the distributions are correlated and
a direction for the informational interaction can be inferred.

 As an example we can imagine how  geographical segregation \cite{4} acts in a large city with a large social or ethnic diversity, such as New York, London, or Paris. In those cities more or less closed communities based on social or ethnic diversity form. Then we could be interested on how those communities interact with each other and which is the topology of interactions  between  them. To understand those interactions we can consider some interesting attributes that are proper of the elements of the different communities and measure the spread of those attributes between them, as wealth, habits, food consumed, etc. In other words we can estimate the information shared between the different communities of a given  sample and establish a link between two communities whenever we recognise an information flow between them. Thus the network of those interactions can give us precious information about the evolution of the mesoscopic systems  defined by the different urban areas inside of the city macrosystem.

In this paper we present a novel methodology based on the Jensen-Shannon divergence \cite{1}  and the Shannon entropy \cite{17} to extract a directed network of  information flows out of a set of populations, where the populations elements are specified by  \emph{n}-dimensional vectors of symbolic attributes. As we already mentioned, this methodology has a wide range of applications from social physics, to economy and biology. We show here an application to genetics,  where we measure the genetic flow between meadows of Poseidonia Oceanica, a Mediterranean seagrass, and an application to semantics, where we measure the semantic flow between different pages of the on-line encyclopedia Wikipedia. In the first case we show that the clusters of the resulting genetic network properly represent the different geographical locations of the meadows, while in the latter case we show that different entries correctly cluster in appropriate semantic categories giving hints to interesting semantic speculations.

\section{Information flow between populations defined in a symbolic attribute space}\label{s2}

 In literature there are different ways to compare probability
distributions \cite{a3}. A convenient one for the kind of systems we want to study is the Jensen-Shannon divergence (JSD
hereafter) \cite{1}. As we better explain  below, we choose it because it is framed in information theory, it takes into account the different sizes of the populations and the
probability distributions don't have to be \emph{absolutely continuous} in each other domini \cite{2}.

Given two probability distributions $P=\{p_1,p_2, ...\}$ and
$Q=\{q_1,q_2,...\}$ of a discrete random variable, the JSD
between $P$ and $Q$ is defined as:
\begin{equation}\label{1}
JSD(P\|Q)\equiv H(\pi_1P+\pi_2Q)-\pi_1H(P)-\pi_2H(Q)
\end{equation}
where $\pi_i$ are weights, that is $\pi_1+\pi_2=1$ and
$H(P)=-\sum_i p_i\ln p_i$ is the Shannon entropy measured in
\emph{nats} \cite{17}.

JSD was introduced in \cite{1} and its properties are well
reviewed in \cite{2}. For our purposes the most important feature
of the JSD is that the two distributions we want to compare
have not to be absolutely continuous in  each other
domini, as it happens for instance in the case of the
\emph{Kullback-Leibler} divergence \cite{1}. In fact we want to
compare distributions of attributes that are not necessarily
shared by all the populations of the system. Moreover the JSD
embeds a weighting system for the different distributions and
it was demonstrated in \cite{2} that the optimal choice for the
weights is the statistical weight of the samples. This feature
is  necessary in order to compare populations that are
different in size. Hence if the number of the elements of the
population defined by the distribution $P$ is $n_1$ and the
number of elements of the population defined by the
distribution $Q$ is $n_2$, we define $\pi_i\equiv
n_i/(n_1+n_2)$.

It has been demonstrated that the square root of JSD  defines a \emph{metric} in the case of populations of the same size, $\pi_1=\pi_2=1/2$, while for different population sizes the
triangular inequality has not been demonstrated yet \cite{bh}. Moreover we have that $0\leq JSD(P\|Q)\leq -\pi_1 \ln
\pi_1-\pi_2 \ln \pi_2 \leq \ln 2$. $JSD(P\|Q)=0\Leftrightarrow
P=Q$ and $JSD(P\|Q)= -\pi_1 \ln \pi_1-\pi_2 \ln \pi_2$ if and
only if $P$ and $Q$ have disjoint domini.

\begin{figure}[!ht]\center
                \includegraphics[width=0.49\textwidth]{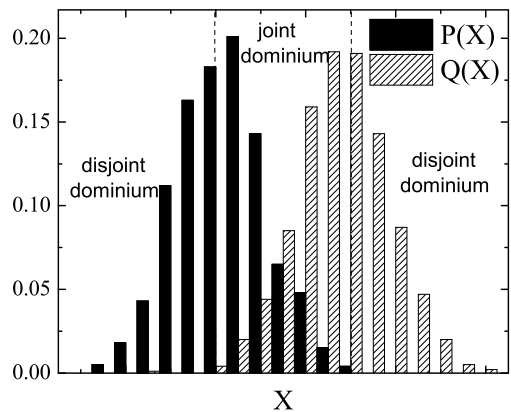}
 \caption{\label{fig1} An example of two probability distributions
 $P(X)$ and $Q(X)$ defined in an attribute dominium $X$ where a fraction $J$
 of the dominium is shared by the two distributions.}
 \end{figure}

JSD measures the information flow between two distributions in
terms of their shared elements and non-shared elements. To
understand the meaning of the JSD we can refer to the example
of the two probability  distributions $P$ and $Q$ defined in a
certain attribute space showed in Fig.\ref{fig1}. $P$ is
defined on an attribute dominium $D_P$, while $Q$ is defined on
a certain attribute dominium $D_Q$. Let us call $X=D_P\bigcup
D_Q$ and suppose that $J=D_P\bigcap D_Q\neq\emptyset$ is the
joint attribute dominium of the two distributions, while
$D=X-J$ is the disjoint attribute dominium of the
distributions. Then Eq.\ref{1} can be split in the two
different domini: $JSD(P\|Q)=JSD(P\|Q)_J+JSD(P\|Q)_D$, where
$JSD(P\|Q)_D=H(\pi_1P+\pi_2Q)_D-\pi_1H(P)_D-\pi_2H(Q)_D=-\pi_1\ln
\pi_1\sum_Dp_i-\pi_2\ln \pi_2\sum_Dq_i$. Then the contribution
given to the JSD by the disjoint domini is a statistical
measure quantifying the non shared attribute distribution
sizes.

For the part of the joint dominium we have that
$JSD(P\|Q)_J=-\sum_J(\pi_1p_i+\pi_2q_i)\ln
(\pi_1p_i+\pi_2q_i)+\pi_1\sum_Jp_i\ln p_i+\pi_2\sum_Jq_i\ln
q_i$. $JSD(P\|Q)_J$ is the entropy of the weighted sum of the
two distributions minus the weighted sum of the entropy of the
distributions, measured in the shared part of the attributes
dominium. From an informational point of view we can say that
if the sum of the distributions is broader than the
 single distributions, this results in a large value of the
divergence. Otherwise if the weighted sum of the distribution
has a larger informative value, hence a smaller entropy than
the one we have from the single distributions, then we obtain a
small divergence from the shared part of the attribute
dominium.

The only issue we get through applying the JSD to a system
composed by many populations is that its maximum value depends
on the population size. That means that we can find cases where
the JSD of two uncorrelated distributions is smaller than the
one of two correlated ones. To avoid this problem we introduce
a new index $D$ defined as the JSD normalised to its maximum
value:
\begin{equation} \label{2}
D(P\|Q)\equiv\frac{JSD(P\|Q)}{-\pi_1 \ln \pi_1-\pi_2 \ln \pi_2}.
\end{equation}
$D(P\|Q)$ has the same properties of $JSD(P\|Q)$ with the
difference that $0\leq D(P\|Q)\leq 1$, where
$D(P\|Q)=0\Leftrightarrow P=Q$ and $D(P\|Q)=1\Leftrightarrow
J=\emptyset$.

\subsection{Directionality}

$D(P\|Q)$ as $JSD(P\|Q)$ is a symmetric quantity in its
arguments, that is $D(P\|Q)=D(Q\|P)$. Hence it doesn't give
information about the directionality of the interaction. In
order to infer a directionality for the information flow we borrow a rationale  from sociology, in
particular from the idea of \emph{geographical segregation}.

Geographical segregation is a concept that is widely used in
many areas of science, such as sociology \cite{4,5},
economy\cite{10}, geography \cite{9}, physics and biology
\cite{32}. It refers to the inequality between population
attribute distributions inside of a metapopulation.
In particular a population inside of a metapopulation is said to be segregated in respect to some attributes if those attributes are found with a consistent probability in that population and are not found with a significative probability in the other populations of the system.
\begin{figure}[!ht]\center
                \includegraphics[width=0.49\textwidth]{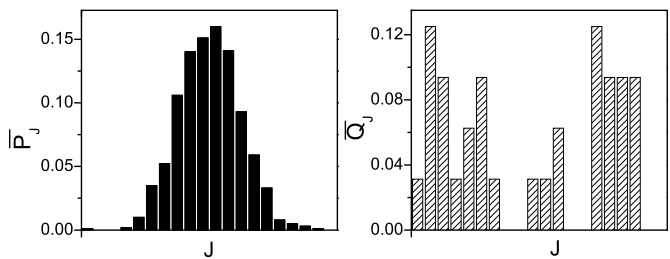}
 \caption{\label{fig2} Following the notation of the example of Fig.\ref{fig1}, here we show the renormalised distribution of the shared elements
  $\overline P_J$ and $\overline Q_J$, in the joint dominium $J$.
  In this particular example it is evident that $\overline P_J$ is a peaked distribution, while $\overline Q_J$ is uniformly distributed.
  Hence the distribution $\overline P_J$ results to be more segregated than $\overline Q_J$.}
 \end{figure}

There are many
indexes in literature to measure geographical segregation
\cite{6}. A popular one is the \emph{Theil's segregation index}
\cite{7,8} and it is based on information theory.
The Theil index is the difference between the total entropy of the system in respect to some attributes and the weighted sum of the entropy of the different populations and it is defined as $T\equiv H_T-\sum_iw_iH_i$, where $H_T$ is the total Shannon entropy of the system, $H_i$ is the entropy of population $i$ and $w_i$ is its statistical weight. If  $T$ is close to 0 it means that those attributes are not segregated in the system, but they are distributed more or less uniformly through it. If $T$ is consistently larger than 0, it means that those attributes are segregated in one or more populations of the system.

The Shannon entropy is a well defined measure to estimate the
amount of inequalities represented by a probability distribution. It is
large when the attribute frequency distribution is uniform and
it increases with population size. In our case a large entropy
for an attribute ensemble represents the fact that different
attributes are equally mixed and it is a hint of small
segregation in the attribute space. Otherwise a small value of
Shannon entropy is associated to a large inequality between
attribute frequencies and to a small number of different
attributes and it is an evidence of  segregation for the
population in the attribute space, where exchanges with other
populations are a few.
 Then, in general terms, if an information flow is detected between two populations we can argue that the origin of the
information  is in the most segregated distribution, where the Shannon entropy is smaller (see Fig.\ref{fig2}).

Hence, given two distributions $P$ and $Q$ between which an information
flow is detected, to infer the directionality of the flow
 we first consider the inequality
of the two distributions in the joint dominium.
To do that we consider the distributions $\overline P_J$ and $\overline Q_J$, that are the distributions of the elements of $P$ and $Q$ that belong to the joint dominium $J$, with their frequencies renormalised to unity in $J$ (see Fig.\ref{fig2}).

The number of attributes shared by two distributions is the
same for both the distributions, hence the entropy $H$ measured
over the joint dominium $J$ depends only on the
relative frequencies of the attributes. In particular more
peaked distributions have smaller entropy than broader
distributions. Then we have to take into account the fact that
the population sizes are different. In particular it is
important to understand which is the ratio of the shared elements within the whole population. To do that
we define  the index $\mu_{P}\equiv
\frac{\sum_J p_i}{\sum_{X} p_i}$, $\mu_{Q}\equiv
\frac{\sum_J q_i}{\sum_{X} q_i}$  and we have $0<\mu_{P,Q}\leq 1$.
 If for a certain distribution $\mu$ is close to
one, it means that the shared attributes
are the dominant part of that sample. Then an estimator
for the information flow directionality between $P$ and $Q$ can
be defined as
\begin{equation}\label{3}
I(P\rightarrow Q)\equiv -\mathrm{sign}\left[\frac{H(\overline P_J)}{\mu_P} -
\frac{H(\overline Q_J)}{\mu_Q}\right].
\end{equation}

If $I(P\rightarrow Q)=+1$ the information carried by the
attributes in the joint dominium of $P$ is larger than the
information carried by the attributes in the joint dominium of
$Q$. Then we can infer an information flow from the attribute
distribution $P$ to $Q$. Otherwise, if $I(P\rightarrow Q)=-1$, we can infer an information flow from the attribute
distribution $Q$ to $P$.

\section{Genetic flow between  seagrass meadows}\label{s3}

In this section we build the genetic flow network within subpopulations of  Posidonia Oceanica (PO hereafter). PO is a seagrass that is endemic to the Mediterranean Sea \cite{a4}. It can reproduce either sexually via floating fruits, either asexually spreading stems, tough the latter way is the most common one.  PO is a determinant species for the Mediterranean ecosystem, since its large colonies give shelter to many other species. Generating a network of directed genetic flows between the meadows helps to understand how this species grew and populated the Mediterranean sea.

The dataset is composed by a  set of $N=37$  meadows of PO, geographically distributed in $N$ points. For every meadow, or population,
a variable number $M=40\pm 5$ of ramets, or individuals, were genotyped  in terms of $n=7$ \emph{microsatellite markers} \cite{a4}.

Microsatellite markers are tandem nucleotide repeats that are present in the non-coding region of DNA \cite{13}.
Their function is not understood yet, but their regularity makes them optimal markers to identify individuals. In fact it is via microsatellite markers that DNA is investigated in forensic trials. The same dataset was already analysed in \cite{a5,a9} with  different genetic distances. Since each allele of a microsatellite marker is characterized by the number of repetitions of a specific DNA motif occurring at that microsatellite locus, each individual or ramet belonging to a given population is characterized by a set of $n=7$ pairs of integer numbers.

An example  of  such a ramet is [(151, 161), (164, 164), (210, 210), (234, 238), (159, 171), (178, 178),   (178, 180)].
The allele number for each locus is expressed as a pair because  PO is a \emph{diploid organism}, that is its DNA is made of two complete sets of chromosomes,
so that each number belongs to a given chromosome. However we don't know the exact order for the numbers of each pair, that is we don't know which are the numbers belonging to a given chromosome and the numbers belonging to the other one. For this reason the allele repetitions  for each locus are ordered by their size.

The classical way to treat this kind of data is to consider the ensemble of alleles at each locus and then to average over the loci \cite{a6}. Nevertheless this  approach gets rid of the correlations between the alleles belonging to different loci.

\begin{figure*}[!ht]\center

                \includegraphics[width=0.85\textwidth]{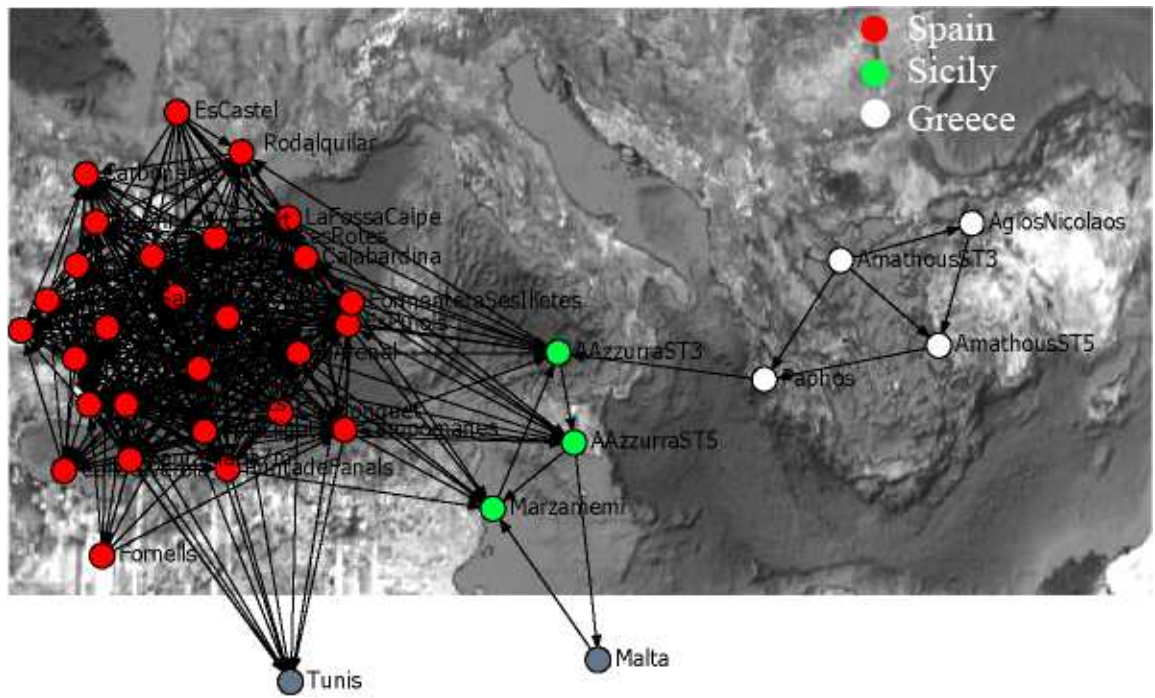}
 \caption{\label{f1} (Color online) Directed genetic flow network of PO meadows in the Mediterranean Sea. The network is displayed via a classical algorithm of spring embedding at its percolation threshold, without additional geographical information. Nevertheless the genetic clusters efficiently reflect the geographic locations for the meadows and the directions of the genetic flows agree with standard evolutionary theories for the PO.}
 \end{figure*}

To avoid this problem we represent each ramet in a 7-dimensional space, $\mathrm{I\!N}^7$, where each dimension is a specific locus. Then for each ramet we consider all the possible combinations of  all the pairs of alleles in the 7 loci. In this way we obtain for each ramet  $2^7=128$ points in the loci space, each point representing an equiprobable \emph{gamete} representative of the ramet.

As an example if we have a ramet with two diploid loci, (125,127) and (400,404), then we can represent that ramet with $2^2=4$ points in a 2-dim space: (125,400), (125,404), (127,400), (127,404).
In this way each population is represented by a set of $5078\pm597$ points in $\mathrm{I\!N}^7$, which is  the statistical sample characterizing the probability distribution function in that space. Moreover every homozygous locus gives birth to two equal points in  $\mathrm{I\!N}^7$, this feature giving statistical strength to homozygosity in the resulting density distribution for the population. Thus we obtain 187904 representative points for the 37 populations.

Each meadow is completely specified by its probability distribution in the 7-dimensional loci space, each point of the distribution giving us the probability that a certain gamete is present in a given meadow and by its size. To generate the network of genetic flow between the meadows we apply Eq.\ref{2} and Eq.\ref{3} to the processed dataset.

The measurement of the directional genetic flow between
meadows gives us a list of all the possible pairs of meadows
separated by a directional genetic distance. Then we order the
meadow pairs for increasing values of their genetic distance,
and we define a network of meadows considering two meadows as
linked when their genetic distance is smaller than a given
threshold.

When increasing the value of the threshold we obtain a growing
network where the first links to form are the strongest in a
genetic sense.  We can analyse the network at different
thresholds to see how the different clusters form and merge. A significative threshold to analyse the network
 is the percolation threshold (PT
hereafter), when the main clusters of the network  connects \cite{30,40}.

 In Fig.\ref{f1} we show the resulting network at its percolation threshold. The network is displayed by a classical algorithm of spring embedding
 \cite{a7} that shows the emerging clusters. No geographical data are considered to
 draw the network and the geographical map in the background is given to have an idea of the geographical spaces that are involved.
Different colors are given to meadows belonging to different geographical areas. As we can see the algorithms presented in Sec.\ref{s2} efficiently split the geographical clusters of Spain, Sicily and Greece. In particular the genetic channel between East Mediterranean and West Mediterranean Sea is well recognised with the link between a Greek meadow and a Sicilian one. Moreover the detected direction of this latter genetic flow is in agreement with evolutionary hypothesis for the spread of PO in the Mediterranean Sea \cite{a5}.

%

\section{Semantic flow between Wikipedia pages}
\begin{figure*}[!ht]\center
                \includegraphics[width=1.\textwidth]{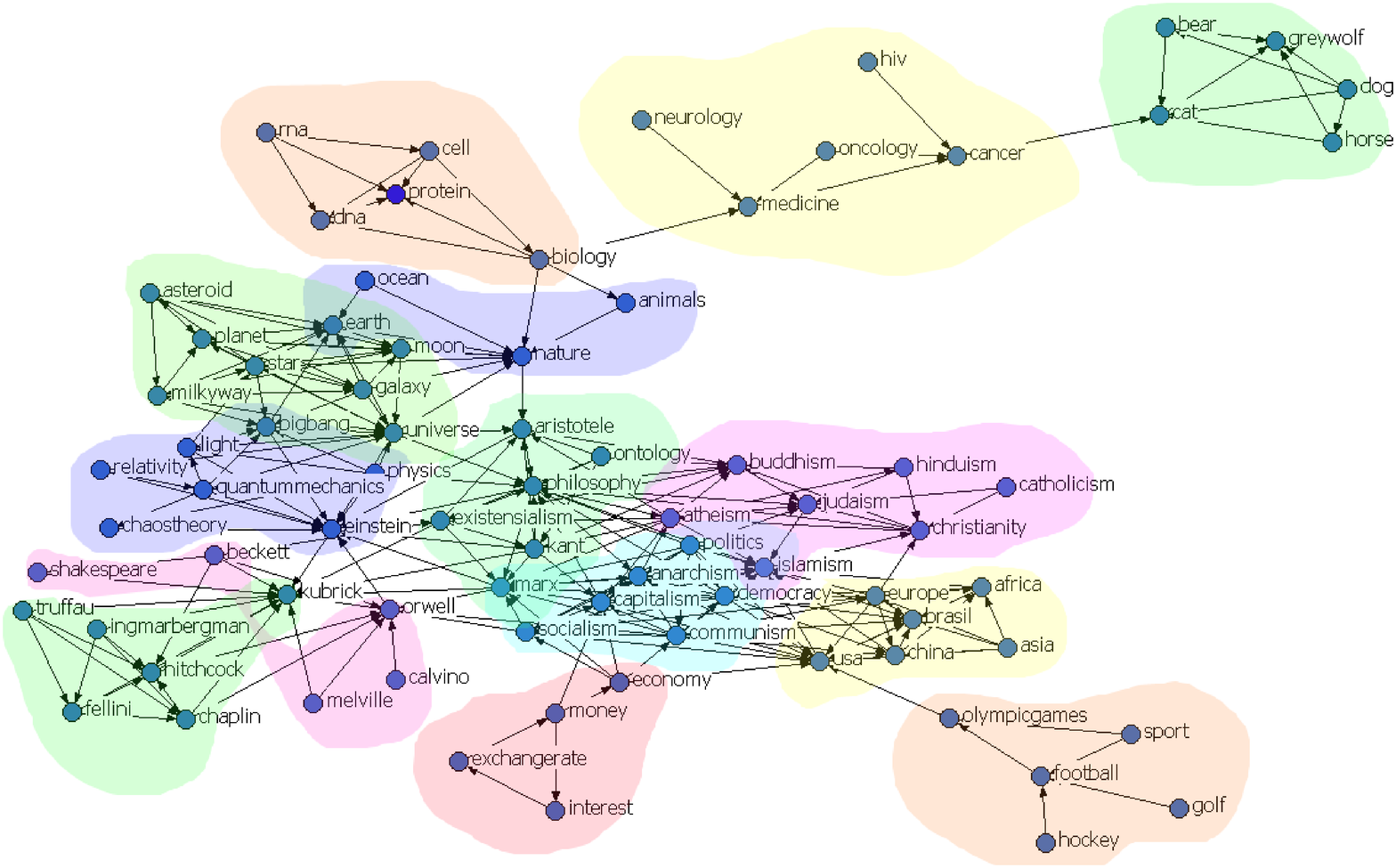}
 \caption{\label{f2} (Color online) Semantic flow network between 78 Wikipedia pages selected within 14 different categories. The network is displayed at the percolation threshold via an automatic spring embedding technique. The clusters efficiently split the different semantic areas.}
 \end{figure*}

In this section  we show an application of the method presented in Sec.\ref{s2} for the detection of semantic flows between different written texts. In what follows we measure the semantic flow between a set of selected Wikipedia pages. In this paper we are mostly interested to show the reliability of the method, but its application to semantics can lead to interesting research for automatic semantic classification in digital media, or for human machine interfaces \cite{34}.

We consider 78 entries of the Wikipedia, selected between 14 different categories and we calculate the directional semantic flows between each pair of pages. First of all we process the text to get rid of all its structural tokens, as articles, punctuation, the most common adverbs and adjectives. After that we lemmatise the text, that is we reduce all the verbs to their infinitive forms and all the plural words to singular.

Then we consider each page as a population where the elements are the different processed words forming  a 1-dimensional attribute space.
Hence each page is defined by its content word distribution and by its size.

As in  Sec.\ref{s3}  we apply Eq.\ref{2} and Eq.\ref{3} to our dataset, thus obtaining a list of all the pages separated by a semantic distance. We order the list for increasing values of $D$ so that we can analyse the network at different thresholds. Again an interesting threshold is the percolation threshold, since it shows the active semantic channels between very different areas of knowledge.

The resulting network is displayed in Fig.\ref{f2}. Again the introduced algorithm is able to efficiently split the pages belonging to different categories in different clusters.
Moreover we can see how the semantic flows can delimit different areas of knowledge.

Many semantic aspects emerge from this  analysis.
For instance it is interesting to notice how the semantics used to describe movie directors is interlaced with the one used for literature writers.
 It is interesting to see how the semantics describing Karl Marx forms a bridge between the one used for philosophy and the one used for  socialism.
 It is interesting to see how the economics semantics is common to all the politics, while the financial one is proper only of capitalism.
  It is interesting to see how the different countries have a semantic description that lies between the politics one and religions one and how philosophy forms a semantic channel between politics and science.

\section{Conclusions}

In this research we  introduced a  method to measure directional information flow between different populations belonging to a given system. The definition of such a system is very general, so that the applicability of the method is wide, even if the relation with metapopulation dynamics is evident \cite{a11}. In particular the elements of the system can be described by a multidimensional vector of either numeric or symbolic attributes and the method takes in account the different population sizes.

The improvement of this methodology over the classical ones used to compare probability distributions \cite{1} is that it is designed for a many-populations system, giving the chance to build a network of information flow. Moreover the application of ideas coming out of geographical segregation studies allows to address the question of directionality in the interaction, transforming the static idea of correlations or divergence between probability distributions, in a dynamical idea of information flows between subsystems of a given macrosystem.

We showed two simple applications of the method to different scientific fields as   semantics and genetics. In the first case  we showed how the method can recognise the
geographical locations of seagrass meadows via microsatellite markers. In the latter case we showed how the method can easily map a portion of the semantic space via the analysis of word distribution in Wikipedia entries.

The topic of genetic distance is wide and it is possible to find many different approaches to the problem in literature \cite{a10}.
To review them in comparison to our measure is out of the purpose of this paper. In particular the interested reader can refer to \cite{a12} for a more detailed analysis. For now we can say in very general terms that the major improvements that our method brings in respect to other genetic distances resides first of all in the fact that our measure accounts of different population sizes and infers a directionality for the interaction. Another novelty of the approach is to consider the multidimensional gamete space, instead of considering the  allele abundance averaged on the different loci. Thus considering the correlations between alleles in different loci narrows the analysis in a more recent evolutionary time-scale.

The  analysis of the whole semantic space as represented by the network of semantic flows between the entries of the whole Wikipedia has non trivial properties and  is presented in \cite{a8}.

We think that apart from the mentioned case studies the presented methodology can have important applications in fields such as sociology, sociogeography and economics.

\textbf{Acknowledgments} Supported by Ministerio de Ciencia e Innovaci\'on and Fondo Europeo de Desarrollo Regional through
project FISICOS (FIS2007–60327). Many thanks to F. Alberto and E. Serr\~ao for the useful discussions and to A. Herrada for its patient lessons on molecular biology.

\end{document}